\documentclass[twocolumn,showpacs,preprintnumbers,amsmath,amssymb]{revtex4}

\usepackage{graphicx}
\usepackage{dcolumn}
\usepackage{bm}

\begin{document}

\preprint{}

\title{The Higgs Mechanism and The Vacuum Energy Density Problem}

\author{Alp Deniz \"Ozer}
\email{oezer@theorie.physik.uni-muenchen.de}
\affiliation{%
Ludwig Maximillians University  Physics Section, Theresienstr. 37,
80333  Munich Germany \\
}%

\date{\today}

\begin{abstract}
We discuss the vacuum energy density term resulting from the
spontaneous breakdown of the  electroweak gauge symmetry, in the
Higgs Mechanism. We alternatively expand  the scalar field at one
of the degenerate states that lie outside the circle of minimum,
such that the Higgs Potential becomes free of any constant field
term, and describes a true vacuum. We show that this true vacuum
requires  a slightly smaller quartic coupling, if the same $v$ and
$m_H$ values of the electroweak model are imposed. We propose that
this small difference ( exactly 20 percent ) can be utilized as a
test to distinguish and identify the true vacuum, in future
experiments at LHC. We shortly discuss the resulting new Higgs
Potential, and its cosmological implications.
\end{abstract}


\maketitle

\section{Introduction}
The Higgs Mechanism~\cite{Higgs:1964ia}\cite{Higgs:1964pj}
implemented in the Electroweak Model~\cite{Weinberg:1967tq}
\cite{Salam:1968rm} \cite{Glashow:1970gm} is so far the only tool
that can generate masses for gauge bosons and
fermions~\cite{'tHooft:1972fi}. However it predicts  the existence
of a massive Higgs boson that still awaits experimental
confirmation. Yet another pending problem in the Higgs Mechanism
is that it generates a constant term which contributes to the
energy density of the vacuum. The magnitude of the contribution
can be estimated. Thus a Higgs mass slightly above the current
lower bound, leads to an energy density roughly $55$ orders of
magnitude larger than the current value. It is very hard to
account for such a tremendous mismatch between theory and
observation. In this paper we argue that this contribution might
follow from our miss understanding of the mechanism. A brief
review of the Higgs Mechanism is therefore essential for our
discussion. We start with the lagrangian of the spinless scalar
field
\begin{equation}
L_\phi=(D_{\mu} \phi)^\dagger (D^{\mu} \phi)+\mu^2 \left(
\phi^\dagger \phi\right) -\lambda
 \, \left( \phi^\dagger \phi\right)^2
\end{equation}
where $\phi$ is a doublet of  complex scalar fields transforming
under the $SU(2) \times U(1)_Y$ electroweak gauge symmetry and
$D_{\mu}$ is the gauge covariant derivative.
\begin{equation}
\phi =  \left( \begin{array}{c}
  \phi^+ \\
  \phi^0 \\
\end{array} \right)
\end{equation}
The electric charges of the components in $\phi$ follow from $Q =
I_3+Y/2 $, and the components are made up of $4$ real valued
scalar fields
\begin{equation}\label{normalization}
  \phi^+ = \frac{\phi_1 + i \phi_2}{\sqrt{2}} \ \ \ \ , \ \ \ \ \
  \phi^0 = \frac{\phi_3 + i \phi_4}{\sqrt{2}}
\end{equation}
The potential part of the Higgs lagrangian has a minimum when
$\mu^2>0$ and $\lambda >0$.
\begin{equation}\label{scalar-pot}
V(\phi) = -\mu^2 (\phi^\dagger \phi) + \lambda (\phi^\dagger
\phi)^2
\end{equation}
The minima follows from the functional derivative $\partial V /
\partial \phi=0$. This gives us the condition : $\phi^\dagger
\phi = \mu^2/2 \lambda$. A particular solution of the minimum
denoted with $\phi_0$ is
\begin{equation}\label{min}
\phi_1=\phi_2=\phi_4=0 , \, \phi^2_3 = \frac{\mu^2}{\lambda}= v^2
, \,
\phi_0=\left( \begin{array}{c}  0 \\   \frac{v}{\sqrt{2}} \\
\end{array} \right)
\end{equation}
The scalar field $\phi$ could be expanded around the minimum
$\phi_0$ with new field variables $\xi$ and $H$, the former fields
give rise to massless 'would-be' goldstone
bosons~\cite{Goldstone:1962es} \cite{Goldstone:1961eq} and the
latter is the massive Higgs field. A useful expression for the
scalar field expanded around the minimum that eliminates the $\xi$
fields is
\begin{equation}
\phi(x) \approx \phi_0(x)= e^{- i \, \sigma \cdot \xi(x) / v } \ \
\left(
\begin{array}{c}
  0 \\
  \frac{v + H(x)}{\sqrt{2}}\\
\end{array} \right)
\end{equation}
The complex scalar field is thereby reduced solely to the massive
Higgs field $H$ by means of the unitary
gauge~\cite{'tHooft:1972fi}. If the above expression for $\phi(x)$
is substituted back in $V(\phi)$ we find the Higgs potential
around the minimum
\begin{equation}\label{higgs-pot}
\begin{split}
 V(H) & =  \frac{1}{2}m^2_H  H^2  \pm  \sqrt{\lambda} \mu  H^3 +
  \frac{\lambda }{4}H^4 -\frac{\mu^4 }{4\lambda} \\
 \end{split}
\end{equation}
Herein the first term gives us the mass of the Higgs particle and
the last term is a constant, contributing to the field. We have
\begin{equation}\label{higgs-parameter}
m^2_H = 2\mu^2 > 0 , \,  v = \pm \frac{\mu}{\sqrt{\lambda}}, \,
V(0) = -\frac{\mu^4}{4 \lambda}= -\frac{1}{8}m^2_H v^2
\end{equation}
We have deliberately kept the $\pm$ sign in $v$. It affects only
the term with $H^3$ in $V(H)$, other terms  do not suffer any
change in sign. Finally, $V(0)$ is equal to $V(H)$ at $H=0$. The
Higgs potential has the following feature at $H=0$.
\begin{equation}\label{charI}
\begin{split}
 &\left( \frac{\partial V}{ \partial H} \right)_{0}   =  0 \\
 &\left(  \frac{\partial^2 V}{ \partial H^2} \right)_{0} =  2 \mu^2  = m^2_H >
 0 \\
\end{split}
\end{equation}

\section{Discussion about the Vacuum}
As we wanted to expand the scalar field around the minimum of the
potential $V(\phi)$, we informidably assigned the scalar field the
value $\phi=\phi_0$. The potential $V(\phi)$ at the minimum
assumes the value
\begin{equation}
V(\phi_0) = -\frac{\mu^4 }{4\lambda} = -\frac{1}{8}m^2_H v^2
\end{equation}
and has dimensions of energy density. A rough estimate of its
value gives
\begin{equation}
  -\frac{1}{8} m^2_H v^2  \approx -2 \cdot 10^8
    \text{GeV}^4
\end{equation}
Here $m_H$ is chosen to be approximately $150$ GeV, and $v=246$
GeV. There is so far $no$ way to get rid of this term\footnote{It
is frequently argued in literature that adding a constant term to
the lagrangian, which in principle {\it could}  remove the term,
is {\it not} a remedy. We subscribe to this point of view, since
it is not natural.}, therefore it should somehow manifest itself
in our universe. Direct calculations from cosmology, give us a
vacuum energy density that is roughly 55 orders
smaller~\cite{Groom:2000in}\cite{Eidelman:2004wy}. We could say
that the vacuum energy density is practically zero. As a result
the above term is obviously indicating that there is something
wrong with either the Higgs Mechanism, or the way we make use of
it\footnote{Another possibility, is to consider the spontaneous
breakdown of the symmetry within the dynamics of an expanding
universe\cite{Kim:2002fd}, where the driving agent is the vacuum
energy. Even if this were the case it would be impossible to
understand why the expansion rate is $currently$ so low for such a
huge constant term, because it should be noted here that  Higgs
potential in eq.(\ref{higgs-pot}) describes, at $H\approx0$ the
{\it current} vacuum energy of our universe, and not that in the
{\it past}. The current cosmological constant is not in favor of
this huge energy. A zero vacuum energy, that would confirm the low
expansion rate, could be obtained at $H=\mu/\sqrt{\lambda}$, ( for
$v=-\mu/\sqrt{\lambda}$ ), but this identically washes all mass
terms in the standard model lagrangian out.}. Note that there is a
negative sign appearing in the term, independent of the sign of
$v$. Since energy should always be positive ,the term also fails
to describe a well defined physical quantity. Therefore our
interpretations is that the term basically indicates the shift
between the two levels
\begin{equation}
   V(\phi=0) \leftrightarrow V(\phi=\phi_0)
\end{equation}
where $V(\phi)$ is the scalar potential in eq. (\ref{scalar-pot}).

\section{Modifying the Higgs Mechanism}

We investigate the symmetry breaking procedure where the scalar
field is expanded, not around its minimum but around a state which
satisfies the following condition
\begin{equation}\label{condition}
    V(\phi_0) = 0 , \ \ \ \   \phi_0 \neq 0
\end{equation}
Actually the above condition tells us that the scalar field should
gain some expectation value without {\it shifting} the Vacuum. The
usual assignment of the vacuum expectation value in the Higgs
mechanism, feels itself free in doing so, and does not respect the
vacuum. Therefore we end up with a huge energy density {\it
missing} in the vacuum. The possible states $\phi_0$  that satisfy
the above condition, lie on a circle, but do not correspond to the
minimum of the scalar potential. It will be shown that this choice
is not arbitrary and has measurable effects, when quartic Higgs
interactions are considered. A discussion is postponed to the end
of this section after the results are obtained.

The expectation value that satisfies the above condition is then
found from  $\phi^\dagger_0 \phi_0 = \mu^2/\lambda$. A particular
solution that satisfies $\phi_0$ is
\begin{equation}
\phi_1=\phi_2=\phi_4=0 , \,  \phi^2_3 = \frac{2\mu^2}{\lambda}=
v^2 , \, \phi_0=\left(
\begin{array}{c}
  0 \\
   \frac{v}{\sqrt{2}} \\
\end{array} \right)
\end{equation}
The scalar field can be expanded around this particular $\phi_0$
by parameterizing it with $\xi$ fields and the $H$ field. The
$\xi$ fields can be removed in the unitary gauge as before through
\begin{equation}
\phi(x) \approx \phi_0(x)= e^{- i \, \sigma \cdot \xi(x) / v } \ \
\left(
\begin{array}{c}
  0 \\
  \frac{v + H(x)}{\sqrt{2}}\\
\end{array} \right)
\end{equation}
By substituting the Higgs field in the scalar potential we obtain
\begin{equation}\label{higgs-mod}
\begin{split}
 V(H) & =  \pm \frac{\sqrt{2}\mu^3}{\sqrt{\lambda}} H + \frac{5}{2} \mu^2  H^2  \pm \mu\sqrt{2\lambda}  H^3
 +  \frac{\lambda}{4}H^4  \\
 \end{split}
\end{equation}
We found out that the expression for the Higgs mass is changed,
the energy density term disappeared and we have a new term
proportional to $H$, which signifies the production of the Higgs
Boson and was not present in the former Higgs potential. We have
then
\begin{equation}\label{modified}
m^2_H= 5\, \mu^2 > 0 , \ \ \ v = \pm \frac{\sqrt{2}
\mu}{\sqrt{\lambda}}, \ \ \ V(0) = 0
\end{equation}
For the Higgs potential $V(H)$ , the following two conditions
occur
\begin{equation}\label{charII}
\begin{split}
 &\left( \frac{\partial V}{ \partial H} \right)_{0}  =   \pm \frac{\sqrt{2}\mu^3}{\sqrt{\lambda}} = \frac{m^2_H
 \,v}{5} \neq 0 \\
 &\left(  \frac{\partial^2 V}{ \partial H^2} \right)_{0} =   + 5 \mu^2  = m^2_H >
 0 \\
\end{split}
\end{equation}
In comparison to the usual Higgs potential in eq.(\ref{charI}), it
is seen that the first term above is differently non zero whereas
the second term is the same.

At this stage it would be relevant to consider perturbations
around the vacuum described by eq.(\ref{higgs-mod}). If we
consider the system as made of a one dimensional $classical$
potential such that
\begin{equation}
\begin{split}
    V  = & \, a\,x+ b \,x^2+c \,x^3+d \,x^4 \ \ , \\
    a = & \, \frac{m_H^2 v}{5} , \ \  b =  \frac{m_H^2}{2} ,\ \  c=v
   \lambda ,\ \  d = \frac{\lambda}{4}
\end{split}
\end{equation}
then oscillations would take place on top of a constant
amplitude\footnote{Here $x_0$ is the constant amplitude resulting
from the inhomogeneous part. We show here that $x_0$ is
classically at the order of the vacuum expectation value $v$.}
$x_0=\frac{a}{2b}\approx O(v)$, for small $\lambda$. As a result
the ground state of the modified Higgs potential in
eq.(\ref{higgs-mod}) contains the information that the vacuum
intrinsically has an expectation value  at the order of $v$, and
perturbations take place on top of it. In contrast, a similar
classical approach for the usual higgs potential in
eq.(\ref{higgs-pot}), which describes a classical potential with
$a=0$, would give instead $x_0=0$, because it lacks of a first
order term in $H$. Thereby it is seen that the ground state of the
usual Higgs potential does not reflect the vacuum expectation
value (vev) i.e., perturbations occur around a vacuum state that
appears to be void of any vev.

From the other side a 1-loop renormalization of {\it both} Higgs
potentials will lead to a relatively small shift in $v$, which
arises from the contributions of the tadpole
diagrams~\cite{Cheng}. Except for the divergent $-\frac{\mu^4
}{4\lambda}$ term in the usual Higgs potential, the
renormalization of the modified Higgs potential will differ from
the renormalization of the usual Higgs potential only due to the
tadpole linear in $H$, and is harmless in the $\phi^4$ theory.

If we assume that $v$ and $m_H$ are both in the {\it modified} and
in the {\it usual} Higgs mechanism the same, then the respective
quartic couplings $\lambda_M$ and $\lambda_U$ will differ in
magnitude from each other, where the subscripts $U$ and $M$ ( M
stands for modified and U for usual ) denote the respective
parameters in eq.(\ref{higgs-parameter}) and eq.(\ref{modified}).
The Higgs mass in the usual and modified potential turns out to be
$m_H = \sqrt{2 \lambda_U} v$ and $m_H = \sqrt{\frac{5}{2}
\lambda_M} v $ respectively. The value of $v$ is fixed through the
Fermi Constant. A comparison of the quartic couplings without
knowing the Higgs mass $m_H$ is possible, eliminating $m_H$ thus
yields
\begin{equation}
    \lambda_{M} = \frac{4}{5}\lambda_{U}
\end{equation}
As a result our condition in eq.(\ref{condition})  predicts a
slightly smaller quartic coupling, for the same $v$ and $m_H$. If
future experiments at LHC find out a deviation from $\lambda_{U}$
in the electroweak model, precisely matching the above deficiency
in the ratio, then this might be attributed to the condition in
eq.(\ref{condition}).

\section{Conclusion}
In the state of the art we got rid of the energy density term by
introducing the above condition, and got confronted with a linear
term in $H$, which explicitly shows the creation of a massive
scalar boson, that was not present before.

The modification that we imposed on the Higgs mechanism might be
understood as a dynamical constraint on the vacuum, which
prohibits the emergence of a shift term, and allows classical
oscillations occur around the true vacuum. By {\it true} we
describe a vacuum that has spontaneously received an expectation
value $v$. This true vacuum meets the predictions of current
cosmology which tells us that  "vacuum" is free of any sizable
energy content and leads to a vanishing cosmological constant. We
proposed a test by means of measuring the quartic coupling. This
test can justify whether the true vacuum and the ground state in
nature is really described as in the modified Higgs mechanism or
not.

We know that the energy density term in the Higgs mechanism is
divergent~\cite{Pes} and doesn't influence observables, but there
is no principle that prohibits it from coupling to
gravity~\cite{Eidelman:2004wy} despite of its negative sign. The
condition in eq.(\ref{condition}) naturally hinders this
catastrophe.

Finally, how the vacuum evolved from the unstable configuration to
its current state which we described by eq.(\ref{condition}) is
still not well understood and studied in models of
inflation~\cite{Eidelman:2004wy}.

\end{document}